\documentclass[runningheads,a4paper]{llncs}

\usepackage{amssymb}
\usepackage{graphicx}
\usepackage{subfigure}
\usepackage{subfloat}
\usepackage{longtable}
\usepackage[all]{xy}

\usepackage{url}
\urldef{\mailsa}\path|{alfred.hofmann, ursula.barth, ingrid.haas, frank.holzwarth,|
\urldef{\mailsb}\path|anna.kramer, leonie.kunz, christine.reiss, nicole.sator,|
\urldef{\mailsc}\path|erika.siebert-cole, peter.strasser, lncs}@springer.com|    
\newcommand{\keywords}[1]{\par\addvspace\baselineskip
\noindent\keywordname\enspace\ignorespaces#1}

\begin{document}

\mainmatter  

\title{Patterns in the occupational mobility network of the higher education graduates.\\
     Comparative study in 12 EU countries}

\titlerunning{Patterns in the OMN of the higher education graduates}

%
%
\author{Eliza-Olivia Lungu%
\and Ana-Maria Zamfir\and Cristina Mocanu}
\authorrunning{Patterns in the OMN of the higher education graduates}

\institute{National Research Institute for Labour and Social Protection,\\
6-8 Povernei Street, Sector 1, Bucharest, 010643 Romania\\
eliza.olivia.lungu@gmail.com,
anazamfir@incsmps.ro, mocanu@incsmps.ro}

%
%

\toctitle{Lecture Notes in Computer Science}
\tocauthor{Authors' Instructions}
\maketitle

\begin{abstract}
  The article investigates the properties of the occupational mobility
network (OMN) in 12 EU countries. Using REFLEX database we construct for each country an empirical OMN
that reflects the job movements of the university graduates, during the first five years
after graduation (1999 - 2005). The nodes are represented by the occupations
coded at 3 digits according to ISCO-88 and the links are weighted with the number
of graduates switching from one occupation to another. We construct the networks
as weighted and directed. 
This comparative study allows us to see what are the common
patterns in the OMN over different EU labor markets. 
\end{abstract}

\section{Introduction}

Job mobility represents an important characteristic of the first years after collage
graduation, as the rate of job change declines with age and experience [1]. According
to human capital theory, youth inexperience lowers the cost of occupational
mobility. This period is called the shopping and thrashing stage and is characterized
by exploration and testing of the labour market.

  Empirical evidences suggest that matching takes place at occupational level as
information obtained by individuals working in a job is used to predict the quality
of the match at other jobs within the same occupation. Thus, those working their
first job are more likely to leave the current job than those working their second,
theird etc. job in the same occupation. 

Investigations on wage formation showed that there are occupation-specific skills
that are transferable across employers. This means that when a worker switches the
employer or the sector, he or she loses less human capital than when changing occupation [2]. 
Higher loss of human capital represents higher costs for mobile workers.
However, some occupations are linked to each other due to the transferability of
skills. Such occupations in which skills and experience can be partially or fully
transferred form career paths. 

By representing the career switches in terms of networks (occupations $\to$ nodes, people changing jobs $\to$ links)
we have first a visual representation of the job mobility and secondly access to all the methods developed lately by the
complex network community to better understand this social and economic phenomena. 

\section{Data}

The empirical occupational mobility network (OMN) is build using REFLEX database
(The Flexible Professional in the Knowledge Society New Demands on Higher Education
in Europe). The database contains information regarding the employment
history of ISCED5 graduates from fifteen countries (Austria, Belgium/Flanders,
Czech Republic, Estonia, Finland, France, Germany, Italy, Japan, the Netherlands,
Norway, Portugal, Spain, Switzerland and UK). For each country it was
drawn a representative sample of individuals which graduated in the academic year 1999/2000, while the data collection took place in 2005. 

We restricted our study to EU countries for which we have data regarding the occupations at 3 digits of the higher education graduates (we excluded Japan, Spain and Switzerland).
We employ the records about their career mobility to built for each EU country, an occupational mobility
network (OMN) that covers their job shifts in the considered period. The generation under study shows a high mobility,
59\% of them changing their job in the considered time period at least once, while 30\% changed it just once. 
 
\section{Occupational Mobility Network}

The first step of the analysis is to generate the occupational mobility network for each country, from the transition matrix between the first and the current job after graduation.
We construct the networks
as directed and weighted. The nodes represent the occupations coded at 3 digits
according to ISCO-88\footnote{The International Standard Classification of Occupations (ISCO-88)
can be consulted at the International Labour Office web page \url{http://laborsta.ilo.org/applv8/data/isco88e.html} (Accessed on October 5, 2012)} 
and the links are weighted with the number of persons moving
from one occupation to another. The occupation change is defined as a modification
in the occupational code at 3 digits of a graduate between the two time moments (first job and current job). We represent as self-loops the job switches in the same occupation.

\begin{figure}[h!]
  \centering
  \includegraphics[width=0.7\textwidth]{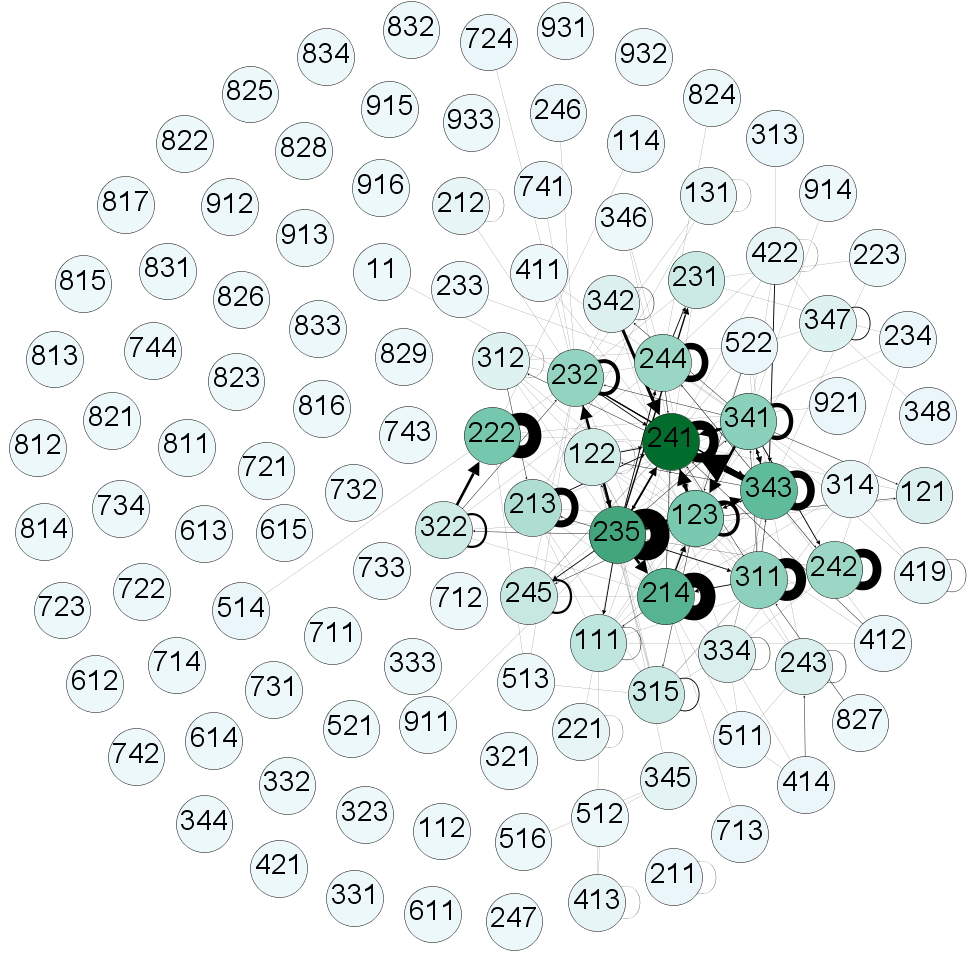}   
  \caption{OMN of Belgium.  
 The thickness of the links is proportional to their weights and the color of the nodes reflects their in-strenght (light green - 
 isolated node, dark green - highly connected node). The labels of the vertexes represent the occupational codes
 according to ISCO-88.}
  \label{fig:g}
\end{figure}

As an example, the occupational mobility network of Belgium is plotted in Fig. \ref{fig:g}. The vertexes are labeled with the occupational codes
and their color reflects the node in-strength (light green - 
isolated node, dark green - highly connected node). The in-strength is calculated at the total weight of the links that point directly to a node $i$ and it is
interpreted as the number of individuals that an occupation i attracted in the considered time span. In the case of Belgium, the node with highest
in-strength centrality is 241 - Business professionals.  
\\
Further we calculate preliminary node specific statistics: node degree (total, in-degree and out-degree), node strength (total, in-strength and out-strength)
and aggregated network statistics: network density, percentage of self-loops and the size of the giant component 
(see Table \ref{tab:tab223}).

The degree of a node $i$ is the number of links connected with it. In the case of directed networks, there are also
defined in-degree, the number of links that point to node $i$ and out-degree, the number of links that leave from node $i$. 
In this context, the in-degree of a node (occupation) $i$ is interpreted as the total 
number of occupations from which it can attract people and the out-degree as the number
of occupations where the labour force can switch from node $i$.  

\begin{table}
\caption{General network properties by country. \small{(122 - Production and operations department managers; 
123 - Other department managers ; 214 - Architects, engineers and related professionals;
223 - Nursing and midwifery professionals; 231 - College, university and higher education teaching professionals; 
235 - Other teaching professionals; 
241 - Business professionals; 
323 - Nursing and midwifery associate professionals; 331 - Primary education teaching associate professionals;
341 - Finance and sales associate professionals; 
343 - Administrative associate professionals;
419 - Other office clerks)}}  
\centering
\resizebox{1\textwidth}{!}{
\begin{tabular}{lccccccccc}
\hline\noalign{\smallskip}
Country & Density & Self loops  & Diameter & Giant & Highest & Highest & Highest & Highest  \\
 &  & (\%)  &   & component & in-degree & out-degree & in-strength & out-strength \\
\noalign{\smallskip}\hline\noalign{\smallskip}
Italy & 0.030 & 11.7  & 5  & 50\% & 341 & 343 & 214 & 214 \\
France & 0.017 & 19.4  & 9   & 54\% & 241 & 343 & 241 & 123  \\
Austria & 0.017 & 13.9  & 6  & 45\% & 241 & 241 & 241 & 241\\
Germany & 0.016 & 13.7  & 7  & 48\% & 241 & 214 & 214 & 214 \\
Netherlands & 0.043 & 9.4  & 6  & 63\% & 241 & 343 & 241 & 123 \\
United Kingdom & 0.030 & 10.6  & 6  & 62\% & 241 & 419 & 241 & 419 \\
Finland & 0.032 & 11.3 &  7  & 59\% & 241 & 343 & 214 & 214 \\
Norway & 0.022 & 13.9 &  8  & 58\% & 122 & 331 & 223 & 323 \\
Czech Republic & 0.035 & 9.5 &  5   & 55\% & 343 & 343 & 241 & 214 \\
Portugal & 0.009 & 25.0 &  7  & 34\% & 231 & 241 & 241 & 241  \\
Belgium & 0.018 & 12.9 &  5  & 52\% & 241 & 235 & 241 & 235 \\
Estonia & 0.020 & 11.9 & 7  & 51\% & 241 & 343 & 241 & 343 \\
\noalign{\smallskip}\hline
\end{tabular}}
\label{tab:tab223}  
\end{table}

For the majority of the countries the occupation that attracts labour force
from the highest number of occupations is 241 - Business professionals. 
In the case of out-degree, the highest value is encounter for 343 - Administrative associate professionals.  
In terms of transferable skills theory, it implies that both occupations use such skills in a high extend. 

The node strength (total, in-strength, out-strength) is calculated in the same manner 
as the node degree, only that this time we sum the weights of the direct neighbors
of a node $i$. As in the case of in-degree, the node with the highest in-strength is
241 - Business professionals, in most of the countries.  

The network density quantifies the proportion of actual links with respect to the all possible ones. The occupational mobility network is sparse
in all the countries, having a low network density (see Table \ref{tab:tab223}). Our result confirms the importance of education in labour market allocation, 
as higher education graduates reach just some of the occupations, especially those requiring higher qualifications.

We also noticed a correlation between EPL (Employment Protection Legislation) and network's
density, in the sense that the more flexible is the national labour market the more diverse is the occupational 
mobility of the collage graduates.

\subsection{Community Structure}
Communities are blocks of nodes that present dense connections
within them and sparse ones with the other communities. In the
context of OMN, we interpret a network community as group of
occupations that actively exchange labour force between them,
thus share a common set of required skills. 

For this analysis we consider the OMN as binary and undirected
and apply modularity optimization method.
Th basic idea of the algorithm is to maximize the modularity
function Q, defined as the difference between the fraction of edges within
communities and expected fraction of edges in a random network.

We estimate the community structure for each country and after that we
overlap them in order to see the common patterns.
On average we found 5, 6 distinct communities in each country. 

\begin{figure}
  \begin{center}
  \includegraphics[width=1\textwidth]{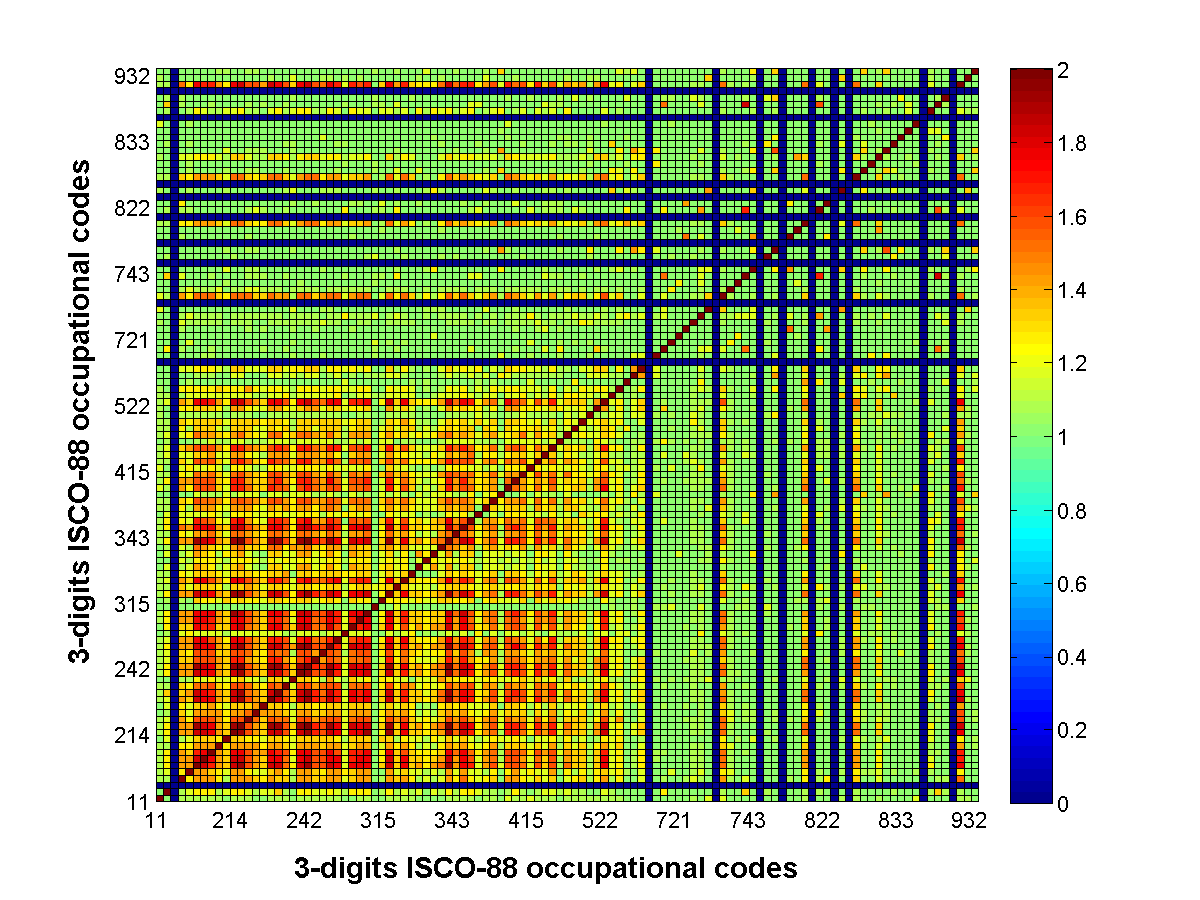}
\caption{Heap map of the similarity index between occupational communities in EU countries (0 - isolated occupation in all the countries and
2 - connected occupation in all the countries).}
  \label{fig:grap_e0}
  \end{center}
  \end{figure}

We calculate an index of similarity between the occupations community structure along the EU countries. The more
countries in which two occupations share the same community, the higher is the value of the index.
We plot the results in a heap map that has on the axis the occupation
codes at 3 digits according to ISCO-88 and in each square the value of the similarity index (Fig. \ref{fig:grap_e0}). 
The redder is the cell, then in more countries two occupations share the same
community. A dark blue cell means that two occupations do not share the
same community in all the countries while a dark red one means that they share the same
community in all the countries. 
So, the blue lines represent isolated occupations (nodes)
over all the countries. We can noticed the red patterns are concentrated in the
first half of the ISCO-88 clasification, which makes sense because here are
the occupations that requires a bachelor degree. We also notice that the occupations from
this area are less connnected with the rest of the network. 

\subsection{3-Node Network Motifs \& Anti-motifs}
\begin{figure}
       \centering
       \includegraphics[width=0.85\textwidth]{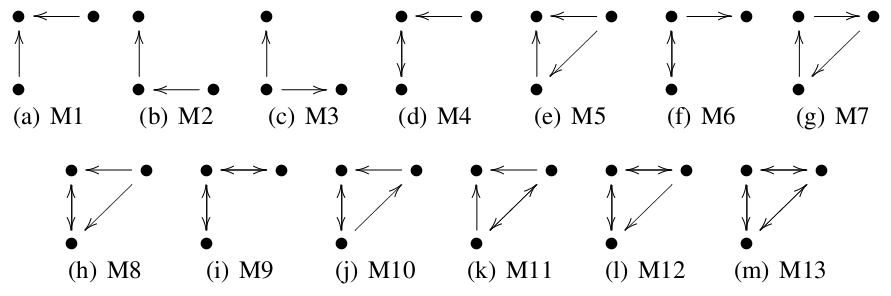}  
 \caption{\small{All types of 3-node network motifs}}
 \label{fig:motifs}
 \end{figure}

The concept of network motifs had been introduced by R. Milo and his collaborators 
to denote ''patterns interconnections occuring in
complex networks at numbers that are significantly higher than those in randomized 
networks''  \cite{5}. 
For this study we are interested in the identification of the statiscally significant three-nodes connected motifs (Fig. \ref{fig:motifs}).
For detecting them we employ Onnela's alghoritm  and compute
the motif intensity $I(g)$ of a subgraph $g$ with links $l_g$:    
\begin{equation} 
\label{eq4}
 I(g)=(\prod_{(ij)\in l_g}{w_{ij}})^{\frac{1}{|l_g|}}
\end{equation}
where $|l_g|$ is the number of links in subgraph $g$.
The total intensity $I_M$ of a three-nodes motif is calculated as the sum of the its subgraph intensities $I_g$.
We preferred this method because we can take into account the weights.

Further, we calculate the motif intensity score in order to test their statistical significance.  
\begin{equation}
 \label{eq6}
\widetilde{z}_M=\frac{I_M-\langle i_M \rangle}{(\langle i_M^2 \rangle -{\langle i_M\rangle}^2)^{1/2}}
\end{equation}
where $i_M$ is the total intensity of motif M in the reference networks. The null-model network is constructed
as an ensemble of random networks generated by reshuffling the empirical weights. \cite{6}.

Table \ref{tab:tab11} summarizes the statistically significat motifs, by country. 
We also include in the table the motifs that are under represented
in the OMN, as anti-motifs. We notice that M3 is poorly represented
in all the countries, beside Norway, which means that there are not occupations
from where the labour force just moves away. Rather, we could say that it
is more a reciprocal exchange between occupations (M12, M13).
\begin{table}
\caption{Motif fingerprint by country (1 - motif, 0 - absent and -1 - anti-motif)}
\label{tab:tab11}     
\centering
\resizebox{0.9\textwidth}{!}{
\begin{tabular}{lccccccccccccc}
\hline\noalign{\smallskip}
Country & M1 & M2 & M3 & M4 & M5 & M6 & M7 & M8 & M9 & M10 & M11 & M12 & M13  \\
\noalign{\smallskip}\hline\noalign{\smallskip}
Italy & -1 & -1 & -1 & 1 & 0 & 0 & 0 & 0 & 0 & 0 & 0 & 0 & 1 \\
France & -1 & -1 & -1 & 0 & -1 & 0 & -1 & 0 & 0 & 0 & 0 & 1 & 0 \\
Austria & -1 & -1 & -1 & 0 & -1 & -1 & -1 & 0 & 0 & 0 & 0 & 0 & 1 \\
Germany & -1 & -1 & -1 & 1 & 0 & 0 & 0 & 0 & 1 & 0 & 0 & 1 & 0 \\
Netherlands & 0 & -1 & -1 & 1 & 0 & 0 & 0 & 0 & 0 & 0 & 0 & 1 & 1 \\
UK & 0 & 0 & -1 & 0 & 0 & 0 & 0 & 0 & 0 & 0 & 0 & 0 & 1 \\
Finland & 0 & -1 & -1 & 0 & 0 & 0 & 0 & 1 & 0 & 0 & 0 & 1 & 1  \\
Norway & 0 & 0 & 0 & 1 & 0 & 0 & 0 & 0 & 0 & 0 & 0 & 1 & 0  \\
Czech Republic & 0 & -1 & -1 & -1 & -1 & 0 & -1 & 0 & 0 & 0 & 0 & 1 & 1  \\
Portugal & 0 & -1 & -1 & 0 & 0 & 0 & 0 & 0 & 0 & 0 & 0 & 1 & 0  \\
Belgium & -1 & -1 & -1 & 0 & 0 & 0 & 0 & 1 & 0 & 0 & 0 & 1 & 1  \\
Estonia & 0 & 0 & -1 & 0 & 0 & 0 & 0 & 0 & 0 & 0 & 0 & 1 & 0  \\
\noalign{\smallskip}\hline
\end{tabular}}
\end{table}

\section{Conclusions}

We investigate patterns of occupational mobility by employing a network based approach in 
which the nodes are occupations are 3 digits and the links are weighted with the flows of individuals moving from one occupation to another. Such an approach helps us to better visualize paths of mobility and calculate network 
indicators in order to understand models of connectivity between different occupations.

\end{document}